\documentclass[conference]{IEEEtran}
\IEEEoverridecommandlockouts
\usepackage{cite}
\usepackage{amsmath,amssymb,amsfonts}
\usepackage{algorithmic}
\usepackage{graphicx}
\usepackage{url} 
\usepackage{textcomp}
\usepackage{xcolor}
\usepackage{balance}
\usepackage{booktabs}
\usepackage[a4paper, total={184mm,239mm}]{geometry}
\def\BibTeX{{\rm B\kern-.05em{\sc i\kern-.025em b}\kern-.08em
    T\kern-.1667em\lower.7ex\hbox{E}\kern-.125emX}}

\AtBeginDocument{%
  \setlength{\abovedisplayskip}{1.0ex plus 2.8pt minus 1.4pt}%
  \setlength{\belowdisplayskip}{1.0ex plus 2.8pt minus 1.4pt}%
  \setlength{\abovedisplayshortskip}{0pt plus 2.8pt}%
  \setlength{\belowdisplayshortskip}{1.0ex plus 2.8pt minus 1.4pt}%
}

\begin{document}

\title{From Layout to Physics: Physics-Grounded \mbox{Modeling} for Label-Efficient Analog Layout Automation
\looseness=-1
}

\makeatletter
\def\@IEEEauthorblockconfadjspace{-1.25em}
\makeatother
\author{
\IEEEauthorblockN{\normalsize Minxing Chu, Xinting Jiang, Ruichen Qi, Junyi Luo, Quan Cheng, Mehdi Saligane}
\IEEEauthorblockA{\footnotesize Brown University, Providence, RI, USA\quad
\{minxing\_chu, xinting\_jiang, ruichen\_qi, junyi\_luo, quan\_cheng, mehdi\_saligane\}@brown.edu}
}

\IEEEaftertitletext{\vspace*{-2.0em}}
\maketitle

\begin{abstract}
Analog layout automation has long been a central goal in electronic design automation. However, learning-based analog layout automation remains difficult to scale. Existing approaches often start from schematic netlists and attempt to learn layout generation or layout-quality prediction, but this forward direction is intrinsically ambiguous: the same netlist can admit many valid layouts, while high-quality schematic-layout-QoR triples are scarce, expensive, and often proprietary. Moreover, circuit-level layout quality is difficult to define without a valid testbench and costly post-layout simulation, making large-scale supervision impractical. This paper proposes L2P, a layout-first, two-stage modeling paradigm for analog layout automation. Instead of treating generated layouts as unique golden labels for netlist-to-layout imitation, we use layouts themselves as deterministic sources of physical supervision. Under a fixed PDK and extraction setup, each layout can be mapped to a unique extracted implementation graph and a parasitic profile, revealing the physical consequences of placement and routing decisions. L2P first learns layout-to-physics relationships from abundant open-source layout and extraction data, and then uses a small amount of golden testbench-labeled data to align these physical quantities with circuit-level quality metrics. Across 12 downstream QoR tasks, L2P reduces simulated-label requirements by up to 96.8\% while matching training-from-scratch performance, reaching or surpassing the performance of full-data forward baselines on matched downstream tasks. L2P further supports bidirectional transfer across distinct design sets and PDKs.   

\end{abstract}

\section{Introduction}
Analog layout remains difficult to automate because layout quality emerges
from tightly coupled electrical, geometric, and process-dependent effects. Unlike digital physical design, where many
implementation constraints are encoded in standardized libraries and mature
implementation flows, analog layout still relies heavily on designer
experience to translate circuit intent into physical structure. Therefore, a general flow
that accepts an arbitrary circuit netlist or design description and
consistently produces a manufacturable, high-quality layout without designer
intervention remains difficult. Recent advances in AI have motivated
models that represent circuits directly and learn either layout
characteristics or layout quality-of-results (QoR). These methods have improved
individual stages of the flow, yet their scalability remains constrained by
scarce and ambiguous design data, as well as expensive, testbench-dependent,
and often proprietary post-layout simulation.
\looseness=-1

Instead of treating analog layout automation as a forward supervised learning
problem, we observe that supervision can come from the layout itself. Once a
layout is fixed, its geometry is no longer ambiguous. Under a fixed PDK,
extraction deck, and tool configuration, the layout produces a reproducible
extracted netlist and parasitic RC graph. Thus, although a netlist can lead to
many layouts, causing label ambiguity in conventional training, each layout
here supplies one reproducible physical target, which provides a deterministic
supervision relation under the chosen extraction setup. The extracted
realization records the consequences of the candidate's placement and routing
decisions without first deciding whether that candidate is optimal for a
particular application. This makes both favorable and unfavorable valid
implementations informative: an extractable candidate that is not selected as
the final layout still provides useful supervision for the physical effects it
induces. Therefore, it provides testbench-independent physical supervision that
can be harvested from open-source layouts, intermediate design products, and
programmatic generators without declaring any one implementation to be the
unique golden answer.
\looseness=-1

Based on this observation, we propose a layout-first, two-stage learning
paradigm \emph{(L2P)}  for analog layout automation.
Stage~I (\emph{L2P-Triad}) learns a reusable physical representation from layout geometry, the extracted implementation netlist, and the
parasitic RC graph as three views of the same realized layout. Stage~II
(\emph{L2P-Calibrated}) specializes the pretrained representation to a target
objective from a small target-domain calibration set. This decomposition
amortizes the learning of physical regularities across target objectives,
designs, and processes. Because corpus-scale pretraining uses
extraction-derived physical evidence rather than simulation-derived QoR,
expensive post-layout simulation is moved out of the large-scale training loop
and reserved for calibration and final verification. L2P is designed to transfer across circuits and processes by learning local
geometry--connectivity--parasitic relations instead of features tied to a
particular circuit or testbench. These relations recur across circuit blocks,
and a small design- or process-specific calibration set re-anchors the
representation to a target circuit and PDK. Therefore, L2P can pretrain on
public or generator-produced layouts and adapt to proprietary target designs
without constructing a proprietary corpus at pretraining scale.
\looseness=-1

\begin{figure*}[t]
  \centering
  \includegraphics[width=0.98\textwidth]{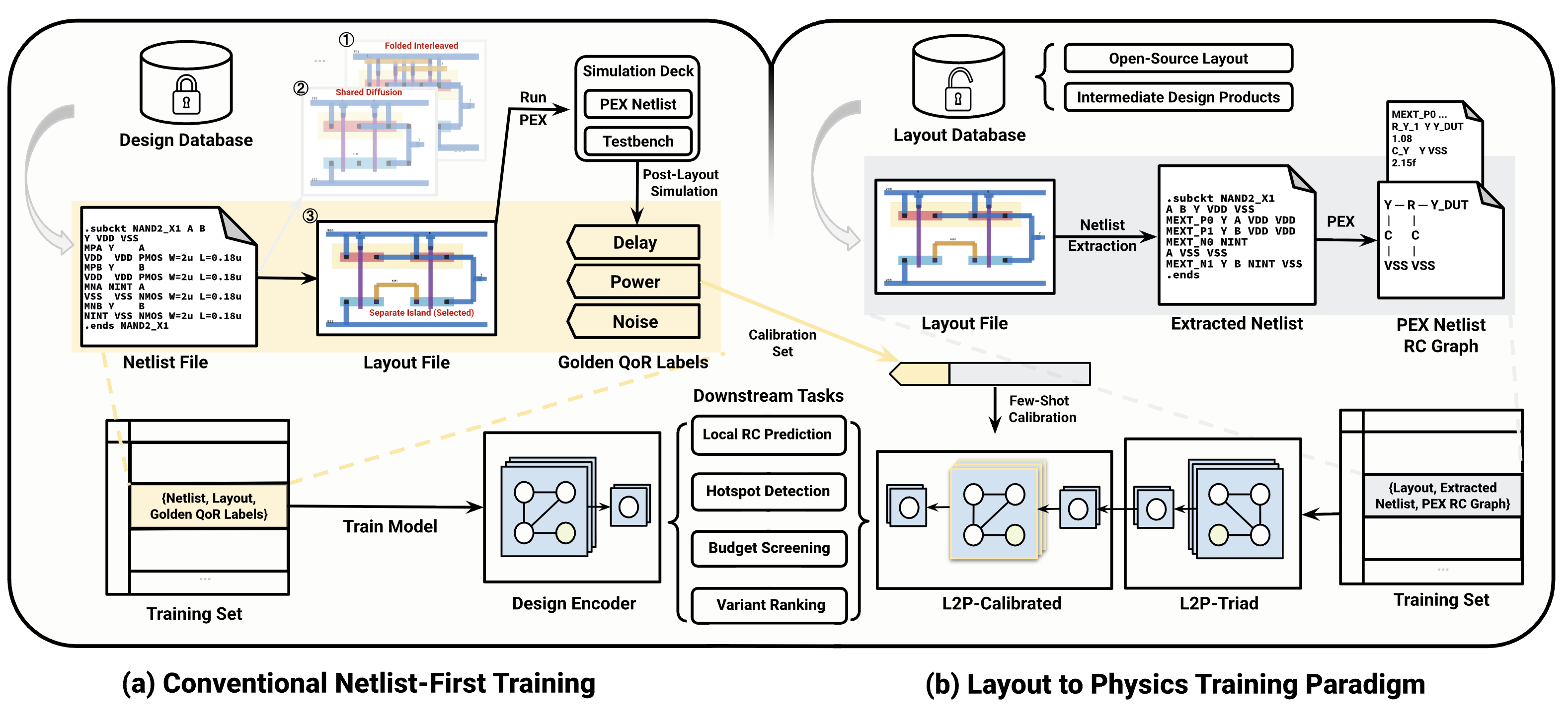}
  \vspace{-12pt}
  \caption{Conventional netlist-first training and the L2P paradigm.}
  \label{fig:l2p-paradigm}
  \vspace{-10pt}
\end{figure*}

Figure~\ref{fig:l2p-paradigm} contrasts conventional netlist-first training
with the L2P paradigm. Conventional pipelines require each layout to be paired
with a testbench-derived QoR label, tying corpus growth to simulation time,
license use, and testbench engineering. L2P instead turns layouts and their
extraction products into corpus-scale physical supervision and reserves golden
labels for a small calibration set. The calibrated representation can then
support heterogeneous downstream objectives.

This work makes three main contributions:
\begin{itemize}
  \setlength{\itemsep}{0.15em}
  \setlength{\parskip}{0pt}
  \setlength{\parsep}{0pt}

  \item \textbf{We propose a novel modeling paradigm} whose supervision signals come directly from the layout itself.

  \item \textbf{We construct a label and simulation efficient two-stage pipeline}, achieving or surpassing matched task performance with as little as 5\% of the simulated labels.

  \item \textbf{We demonstrate lightweight cross-design and cross-process
  transfer}, enabling bidirectional model reuse.
\end{itemize}

\section{Related Work}
\label{sec:related_work}
\vspace{-1pt}

Analog layout automation remains challenging because circuit-specific design
intent is difficult to encode in standardized implementation flows. Nevertheless, prior work has advanced
along two complementary directions: constraint-driven layout synthesis and
learned forward modeling, moving analog implementation toward greater automation
\cite{Gielen2000AnalogCAD,Kunal2019ALIGN,Xu2019MAGICAL}.

Constraint-driven systems encode analog design intent through reusable
primitives, hierarchy, and placement and routing constraints. MAGICAL
integrates constraint inference, placement, and routing, while related work
automates symmetry discovery, primitive construction, common-centroid
layout, and performance-aware routing
\cite{Xu2019MAGICAL,Kunal2020Symmetry,
Madhusudan2021OptimizedPrimitives,Sharma2021CommonCentroid,
Sharma2021PerformanceAwareCC,Chi2021PerformanceRouting}. ALIGN represents an
important system-level milestone in this direction. It accepts a SPICE
netlist, identifies circuit hierarchy and layout constraints, instantiates
parameterized cells, and assembles them into a technology-specific GDSII
layout 
\cite{Kunal2019ALIGN,Dhar2020ALIGNSystem}. These works
demonstrate that substantial portions of the netlist-to-layout flow can be
automated when design constraints are made explicit.
\looseness=-1

A second line of work formulates analog automation as forward modeling:
specifications or transistor-level netlists are mapped to physical-design
decisions or simulation-derived QoR, with training pairs obtained by
executing downstream implementation, extraction, and simulation. Once
trained, these models can amortize expensive analysis or guide exploration
of the design space. Forward circuit modeling has been widely applied to transistor sizing,
topology synthesis, and performance prediction. AutoCkt is an early representative of this formulation. It casts transistor
sizing as a sequential decision problem in which a reinforcement learning
agent adjusts device parameters toward requested specifications,
while circuit simulation evaluates each candidate and provides the learning
signal \cite{Settaluri2020AutoCkt}. GCN-RL makes the input representation
more explicitly structural by encoding the transistor-level netlist as a
graph and incorporating a graph neural network into the
reinforcement learning policy \cite{Wang2020GCNRL}. The graph representation
allows sizing strategies learned from simulated circuit outcomes to transfer
across related circuit topologies and technology nodes. 
CktGNN broadens graph-based forward modeling from sizing a fixed topology to
joint circuit-topology generation and device sizing
\cite{Dong2023CktGNN}. It represents a circuit as a two-level graph
hierarchy constructed from reusable subgraphs, and couples the learned
representation with optimization procedures that search over circuit
topologies and device parameters under simulated performance objectives.
\looseness=-1

Recent work has extended the same forward-learning principle toward physical
implementation. A relational GNN based method encodes
circuit features and positional constraints within a reinforcement learning
policy that generates analog floorplans, which are subsequently completed by
a procedural layout generator \cite{Basso2024AnalogFloorplanning}. TAG pretrains a hierarchical netlist encoder to predict pairwise instance distances measured from manual layouts~\cite{Zhu2022TAG}. FALCON maps circuit topology and device parameters to simulated performance for specification-driven, layout-aware optimization~\cite{Mehradfar2025FALCON}, while Hakhamaneshi et al.~\cite{Hakhamaneshi2022GNNPretraining} pretrain circuit GNNs on inexpensive output-node DC-voltage prediction for few-shot adaptation to unseen metrics and topologies. FALCON's learned forward model
is then used to search for device parameters that satisfy the requested
specifications under a differentiable layout-aware cost.

Although these approaches target different stages, including sizing, topology
synthesis, performance prediction, or floorplanning, they share the same
modeling direction. An early circuit description maps to a later design
decision or a simulation-derived quality metric, and the supervision is
obtained by executing the forward design and evaluation flow.

Despite these achievements, scaling forward supervision remains difficult.
The first issue is that the relation from an initial circuit representation to its physical
implementation is intrinsically one-to-many. The same circuit can admit
multiple valid layouts with different device folding choices, placement
styles, routing topologies, areas, and parasitic tradeoffs. In practice, most available datasets contain only one
or a small number of observed layouts for each circuit. A supervised model is
therefore often trained on a sparse sample of the feasible implementation space,
and its learned distribution is strongly influenced by the designers,
generators, templates, and constraints that produced the available examples.

The second issue is labeling cost: meaningful physical-design labels require
verification, parasitic extraction, and often a circuit-specific post-layout
simulation. The desired supervised example is therefore closer to a
netlist--layout--QoR tuple than to a simple input--output pair. Such tuples
are highly circuit, PDK, and testbench dependent. Moreover, they are expensive to construct at scale. Learned
surrogates can reduce the number of simulations required during later
optimization, but the data used to train those surrogates must still be
obtained from extraction, simulation, or measurements.
\looseness=-1

A particularly relevant response to this label cost is the transfer learning
framework of Liu et al.~\cite{Liu2021FromSpecToSilicon}. In a VCO case
study, a surrogate from parameters to metrics is first trained on comparatively
inexpensive schematic simulations, then adapted with a smaller set of
post-layout simulations from ALIGN-generated layouts and finally calibrated
with silicon measurements. The study demonstrates that a surrogate learned
in a cheaper design domain can be reused with few expensive samples. Together, layout synthesis and transfer learning provide physical
candidates at scale and support label-efficient calibration. However, it is left open how
to learn a physical representation that connects them. Existing transfer surrogates remain circuit-specific, and their calibration
examples still require task-specific simulation or measurement. This motivates
learning from layout-derived evidence before
testbench-specific QoR supervision.
\looseness=-1

\section{The Layout-to-Physics Paradigm}
\label{sec:paradigm}
\vspace{-1pt}

\begingroup
\setlength{\abovedisplayskip}{4pt plus 1pt minus 1pt}
\setlength{\belowdisplayskip}{4pt plus 1pt minus 1pt}
\setlength{\abovedisplayshortskip}{3pt plus 1pt}
\setlength{\belowdisplayshortskip}{4pt plus 1pt minus 1pt}
\setlength{\jot}{1.5pt}

Conventional forward modeling directly learns a mapping from an early design
representation to a task-specific quality-of-result (QoR). Such supervision
couples the learned representation to a particular circuit, testbench, and
performance objective, while requiring every training layout to carry an
expensive Golden label. L2P factorizes this mapping into two stages:
task-independent representation learning from completed layouts, followed by
lightweight adaptation to the QoR of a target task. This learner-agnostic supervision factorization is instantiated with neural and tree-based models in Section~\ref{sec:eval}.

Let \(n\) denote a netlist,
\(\mathcal{L}(n)\) its set of candidate layouts, and \(\xi\) a fixed
extraction configuration. Extraction $\mathcal{E}_\xi$ maps layout $\ell$ to an
implementation graph \(u\) and a full-RC graph \(p\), while simulation $\mathcal{S}_\tau$ under testbench $\tau$
yields its task-specific QoR:
\looseness=-1
\begin{equation}
\label{eq:supervision-hierarchy}
n\mapsto\mathcal{L}(n),\qquad
\ell\xrightarrow{\mathcal{E}_{\xi}}(u,p)
\xrightarrow{\mathcal{S}_{\tau}}\mathbf{y}_{\tau}.
\end{equation}
A conventional forward surrogate compresses this chain into a direct mapping
from an early circuit description to its final QoR, entangling physical
realization with task interpretation. L2P instead exposes the intermediate
state \((u,p)\). Because this state is reproducible for a fixed \(\xi\) and
exists before any target objective is chosen, reusable physical
supervision is provided.

Each completed realization supplies an L2P-Triad
\(\mathcal{T}_i=(\ell_i,u_i,p_i)\). The three entries are not simply
concatenated raw inputs, but corresponding views of the same realization at
progressively more physical abstraction levels. The layout view \(\ell_i\)
records geometry, \(u_i=G_i^{\mathrm{impl}}\) records the implemented devices,
nets, and connectivity without explicit parasitics, and
\(p_i=G_i^{\mathrm{RC}}\) records the extracted parasitic network. We further
write \(\mathbf{r}_i=\psi(p_i)\) for absolute physical targets and compact
summaries derived from the full-RC graph.

Stage I trains a layout-side representation to retain the physical structure
supervised by the other views:
\begin{equation}
\label{eq:stage-one-objective}
\Theta_s=\arg\min_{\Theta}\sum_i
\mathcal{L}_{\mathrm{I}}\!\left(
\rho_{\Theta}(\ell_i);\mathcal{T}_i,\psi(p_i)\right).
\end{equation}
Here, \(\mathcal{L}_{\mathrm{I}}\) uses extraction-derived supervision from \(\mathcal{T}_i\) to supervise the layout representation
\(\rho_{\Theta}(\ell_i)\), with \(\psi(p_i)\) supplying absolute physical
anchors. Neural instantiations combine contrastive alignment, distillation, physical prediction, and reconstruction; tree instantiations use physical-target prediction. Rather than encoding whether
a layout is good for one objective, Stage I learns what circuit structure the
layout physically implements and what parasitic effects its geometry
produces. Both favorable and unfavorable valid layouts remain informative, because although
their final QoR may differ, the geometry--implementation--physics
correspondence remains valid. After Stage-I training on source-domain
Triads, we denote the retained physical representation by
\(\rho_{\Theta_s}\).
\looseness=-1

Because the same resistance or capacitance can affect gain, settling, timing,
or power differently across circuit topologies, operating conditions, and
testbenches, Stage II learns this task-specific interpretation from $m$ labeled examples
$(\mathcal{A}_i^d,\mathbf{y}_{i,\tau}^d)$ in target domain $d$, where
\(\mathcal{A}_i^d\) contains the views admitted by the downstream protocol.
It fits a task head \(H_\phi\) to
\(\rho_{\Theta_s+\Delta\Theta}(\ell_i^d)\), optionally augmented
with protocol-specific features from \(\mathcal{A}_i^d\), with
\(\Delta\Theta=0\) for head-only adaptation or restricted,
regularized updates for lightweight adjustment. In both cases, the geometry--connectivity--physics relation
learned in Stage I serves as a stable prior across designs.
Stage II uses labeled target examples only to map this relation to target QoR,
without target-domain Triad pretraining or repeated Stage-I learning.

The two-stage factorization also changes how supervision cost scales. A
conventional forward surrogate must attach a task-specific simulated outcome
to all \(N\) training layouts whenever a new objective is introduced. L2P
learns the reusable physical relation once and uses only \(m\) simulations to
specialize it to the new objective. Let \(c_E\) and \(c_S\) denote the marginal
costs of extraction and task-specific post-layout simulation, respectively,
and let \(C_{\mathrm{pre}}\) denote the one-time source-domain Stage-I cost. When the learned
representation is reused across \(K\) downstream tasks, its amortized
per-task cost is
\begin{equation}
\label{eq:amortized-cost}
C_{\mathrm{L2P}}^{(K)}
=
\frac{C_{\mathrm{pre}}}{K}+Nc_E+mc_S,
\qquad
C_{\mathrm{forward}}
=
N(c_E+c_S).
\end{equation}
Net cost savings arise when $K(N-m)c_S > C_{\mathrm{pre}}$. For \(m\ll N\), L2P replaces repeated testbench-dependent simulation with
physical supervision that can be reused across designs and
processes. Its advantage grows as the physical representation is
applied to more downstream tasks.

\endgroup

\section{Evaluation}
\label{sec:eval}
\vspace{-1.5pt}

We evaluate L2P through four complementary questions. First, we measure whether extraction provides physical supervision at substantially lower cost
than post-layout simulation. Second, we quantify how much task-specific
simulation can be avoided when the learned representation is adapted to new
QoR objectives. Third, we test whether the same physical knowledge transfers
bidirectionally across independently constructed design sets and across PDKs.
Finally, we compare L2P adaptation with full-data forward
baselines on matched downstream prediction and sizing tasks.
Upon publication, we will release the complete L2P codebase, datasets, pretrained model
weights, training and evaluation scripts, and experiment configurations needed
to reproduce the full reported results.

\vspace{-3pt}
\subsection{L2P Instantiations and Experimental Setup}
\label{sec:setup}
\vspace{-2pt}
We evaluate L2P using representative neural and tree-based instantiations. The neural Cross-Design instantiation uses a layout graph with up to 128 GDS
polygon nodes, directed four-nearest-center edges, and 10-D features for layer/datatype, normalized bounding-box, area/perimeter, and vertex count. The implementation view is a capped 160-node device--net bipartite graph with terminal-derived bidirectional edges and 10-D polarity, degree, terminal-count, and net-role features. The RC view retains the subgraph induced by up to 192
highest-degree nodes, with R/C type and magnitude encoded
in 8-D endpoint features. Each Triad view uses a three-layer GraphSAGE
encoder~\cite{Hamilton2017GraphSAGE}; Stage II retains only the layout encoder and
adds a two-layer QoR head.
\looseness=-1

All L2P instantiations begin with source-domain L2P-Triad learning and reserve
held-out target QoR for evaluation. Cross-Design uses the normalized target
layout graph, while Cross-PDK applies a deterministic canonicalization
\(\Pi_d\) that maps layers to shared semantic roles and normalizes geometry using layout bounds and common physical units. Process metadata comprises log-scaled node size and nominal voltage. Neither setting requires target-side implementation, PEX, or Liberty during inference. Label Saving and Cross-Work reuse implementation,
extraction-derived, and schematic summaries across downstream objectives. Cross-Design adapts Stage-I representation with parameter-efficient or L2-SP updates, while the other settings use fixed tree-based representations. Each registered task or
closely related task bundle uses its corresponding prediction objective, with
identical target splits and evaluation examples for paired comparisons.

\vspace{-3pt}
\subsection{Cost Estimation}
\label{sec:costest}
\vspace{-2pt}

Equation~\eqref{eq:amortized-cost} separates L2P's efficiency into three
factors: the cost ratio \(c_S/c_E\) of simulation to extraction, the calibration
fraction \(m/N\), and amortization across \(K\) downstream tasks. 

The first factor is expected to favor L2P because parasitic extraction (PEX)
constructs a static parasitic network once from layout geometry, whereas
post-layout simulation repeatedly solves the resulting larger circuit equations over operating points, frequencies, and process
corners. This expectation agrees with prior work identifying simulation as the larger post-layout bottleneck than extraction~\cite{Jain2015PostLayout}. We additionally test this premise using open-source extraction and simulation tools, pairing Magic full-RC PEX with ngspice OP, AC, and transient analyses on the same 172 DRC/LVS-clean SKY130 layouts. 

As shown in Figure~\ref{fig:pex-cost}, panel (a) reports paired extraction and simulation runtimes, while panel (b) summarizes the median simulation/PEX ratio by circuit family and R+C order of magnitude; under equal family-by-size weighting, post-layout simulation costs \(29.75\times\) as much as PEX
(95\% source-clustered bootstrap CI: \(27.96\)--\(31.45\times\)), with the
same direction in all 172 layouts. Section~\ref{sec:label-saving} evaluates the calibration fraction \(m/N\)
across 12 tasks. Subsequent experiments test reuse across design
sources, PDKs, and downstream task formulations.
\looseness=-1

\begin{figure}[t]
  \centering
  \includegraphics[width=0.95\columnwidth]{figures/Figure2.jpg}
  \vspace{-6pt}
  \caption{PEX and Post-Layout Simulation Comparison}
  \label{fig:pex-cost}
  \vspace{-12pt}
\end{figure}

\vspace{-3pt}
\subsection{Label Saving}
\label{sec:label-saving}
\vspace{-2pt}

We evaluate how much task-specific post-layout supervision L2P saves by reusing extraction-derived physical information. We compare
L2P, Scratch-only, and a layout-only pretraining baseline for the same downstream QoR tasks. The layout-only baseline matches L2P in source layouts, model capacity, and training budget, with pretraining driven solely by layout geometry. Scratch-only learns the structural-to-QoR mapping directly, whereas the Label-Saving model uses ten independently seeded 256-tree multi-output ExtraTrees teachers trained on 677 L2P-Triads from 135 source groups. Each teacher maps the same 337-D GDS-only descriptor to five
log-transformed PEX summaries: total resistance (\(\Omega\)),
total and coupling capacitance (fF), and resistor and capacitor counts. Across the ten teachers, the mean, standard deviation, minimum,
and maximum for each target form 20 frozen features. All three routes use identical 192-tree predictors, target splits, target transformations, and acquisition orders. All three heads share netlist, metadata, and schematic inputs. Scratch adds raw geometry, whereas both pretrained routes add PEX/full-RC
features and their respective frozen representations.
\looseness=-1

The downstream cohort contains 100 source groups and 1,867 valid layouts. We
measure task-specific supervision in \emph{Golden calls}, with each call
denoting one post-layout simulation of a layout under its circuit-family
testbench. We evaluate budgets of 25--1,000 Golden calls over five
source-grouped folds and ten label-blind acquisition orders per fold. On schematic-relative targets (dB differences or log-magnitude ratios),
source-level utility is $u=(1+\mathrm{RMSE}/\sigma_{\mathrm{train}})^{-1}$,
where $\sigma_{\mathrm{train}}$ is the floored outer-training
standard deviation. We report
two complementary views: equal-task aggregation weights the 12 QoR tasks
equally, while equal-family aggregation balances the five circuit families.
Aggregate curves undergo equal-weight nondecreasing isotonic regression
before normalization by the corresponding fitted Scratch@1,000 endpoint.

\begin{figure}[t]
  \centering
  \includegraphics[width=\columnwidth]{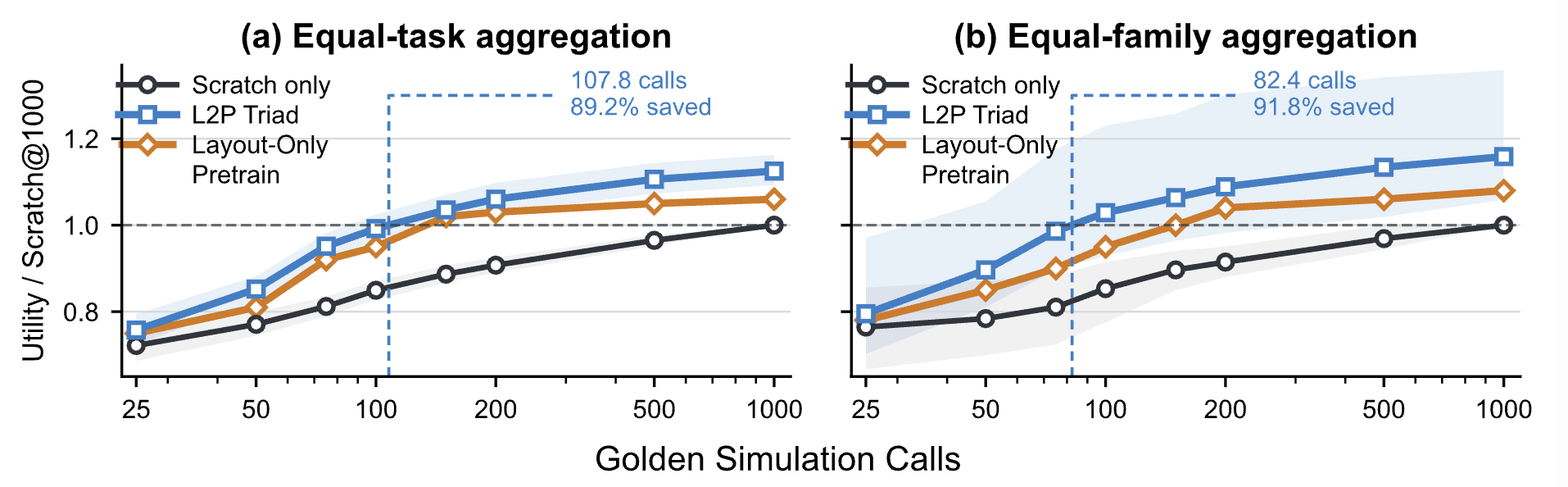}
  \vspace{-2.2em}
  \caption{Golden-call Pareto curves.}
  \label{fig:label-saving-pareto}
  \vspace{-0.8em}
\end{figure}

Figure~\ref{fig:label-saving-pareto} shows that the L2P point-estimate curve
remains above both baselines throughout the evaluated budget range. Table~I reports equivalent budgets obtained by inverting the fitted
L2P curves using linear interpolation in log calls, without extrapolation. L2P matches the
1,000-call Scratch endpoint with 107.8 calls under equal-task aggregation and
82.4 calls under equal-family aggregation, saving 89.2\% and 91.8\%,
respectively. Repeating this procedure in each bootstrap replicate yields paired
95\% intervals of 85.7--91.5\% and 69.6--96.1\%. To isolate the contribution of source-domain pretraining,
we augment Scratch with the same PEX/full-RC inputs.
L2P achieves comparable label savings against this
control, demonstrating that source-pretrained representations
are the primary driver of its label efficiency. Task-level results confirm that the aggregate gains are broadly shared:
L2P saves more than 70\% of Golden calls on 10 of the 12 tasks and more than
90\% on five. The layout-first route matches the endpoint of forward
Scratch-only training with substantially less task-specific post-layout
supervision across both aggregate views and all five circuit families.
\looseness=-1

\begin{table}[t]
  \footnotesize
  \setlength{\tabcolsep}{0.8pt}
  \renewcommand{\arraystretch}{0.98}
  \centering
  \caption{Matched-effect Golden-call savings by scope.}
  \label{tab:label-saving}
  \vspace{-0.85em}

  \footnotesize
  \setlength{\tabcolsep}{0.8pt}
  \renewcommand{\arraystretch}{0.91}
  \begin{tabular*}{\columnwidth}{@{\extracolsep{\fill}}lrrrr@{\extracolsep{0pt}\hspace{-3.5pt}}r@{}}
  \toprule
  Scope &
  \shortstack{Scratch\\calls} &
  \shortstack{L2P\\calls} &
  \shortstack{Calls\\saved} &
  \makebox[2.75em][l]{\hspace*{-2pt}\shortstack{Saved\\(\%)}} &
  \shortstack{Saved vs.\\+PEX/RC} \\
  \midrule

    \textbf{Equal-task macro} &
    \textbf{1000} & \textbf{107.8} & \textbf{892.2} &
    \textbf{89.2\%} & \textbf{84.5\%} \\

    \textbf{Equal-family macro} &
    \textbf{1000} & \textbf{82.4} & \textbf{917.6} &
    \textbf{91.8\%} & \textbf{81.0\%} \\

    \midrule

    \textbf{Buffer macro} &
    \textbf{1000} & \textbf{138.2} & \textbf{861.8} &
    \textbf{86.2\%} & \textbf{84.8\%} \\
    \quad fall time &
    1000 & 216.1 & 783.9 & 78.4\% & 75.7\% \\
    \quad rise time &
    1000 & 78.4 & 921.6 & 92.2\% & 86.3\% \\

    \addlinespace[1pt]

    \textbf{Cascode macro} \textnormal{(settling time, 1\%)} &
    \textbf{1000} & \textbf{756.2} & \textbf{243.8} &
    \textbf{24.4\%} & \textbf{24.4\%} \\

    \addlinespace[1pt]

    \textbf{Common-source macro} &
    \textbf{1000} & \textbf{160.1} & \textbf{839.9} &
    \textbf{84.0\%} & \textbf{82.8\%} \\
    \quad dynamic energy &
    1000 & 32.1 & 967.9 & 96.8\% & 93.7\% \\
    \quad settling time (1\%) &
    1000 & 83.8 & 916.2 & 91.6\% & 79.9\% \\
    \quad absolute slew rate &
    1000 & 406.9 & 593.1 & 59.3\% & 53.0\% \\

    \addlinespace[1pt]

    \textbf{Five-transistor OTA macro} &
    \textbf{1000} & \textbf{121.7} & \textbf{878.3} &
    \textbf{87.8\%} & \textbf{87.1\%} \\
    \quad open-loop gain &
    1000 & 44.7 & 955.3 & 95.5\% & 93.6\% \\
    \quad quiescent current &
    1000 & 191.5 & 808.5 & 80.9\% & 80.7\% \\
    \quad absolute slew rate &
    1000 & 134.4 & 865.6 & 86.6\% & 77.1\% \\
    \quad positive slew rate &
    1000 & 134.4 & 865.6 & 86.6\% & 85.7\% \\
    \quad unity-gain frequency &
    1000 & 289.2 & 710.8 & 71.1\% & 66.9\% \\

    \addlinespace[1pt]

    \textbf{Inverter macro} \textnormal{(dynamic energy)} &
    \textbf{500} & \textbf{26.1} & \textbf{473.9} &
    \textbf{94.8\%} & \textbf{94.4\%} \\

    \bottomrule
  \end{tabular*}
  \par\vspace{1pt}
  \vspace{-18pt}
\end{table}

\vspace{-3pt}
\subsection{Bidirectional Cross-Design Transfer}
\label{sec:cross-design-transfer}
\vspace{-2pt}

We evaluate cross-design transfer between two independently constructed and
complementary public design sets: ALIGN
\cite{Kunal2019ALIGN,Dhar2020ALIGNSystem}, a representative automated
netlist-to-layout framework, and OSIRIS
\cite{Chiari2026OSIRIS}, a scalable back-end dataset in which layout variants are systematically created by exploring transistor design space. The resulting
design sets differ substantially in circuit composition, layout-generation
methodology, and the distribution of physical variations. Bidirectional
transfer between them directly evaluates whether L2P captures
geometry--connectivity--parasitic relationships that remain reusable across
distinct design sources. For each direction, the \emph{transfer model} initializes from source-domain
L2P-Triad pretraining, whereas \emph{target-only} receives no source
pretraining. Model configurations
and source-pretraining budgets are selected only on development folds,
confirmed over five folds and three repeats.

\begin{figure}[t]
  \centering
  \includegraphics[width=0.84\columnwidth]{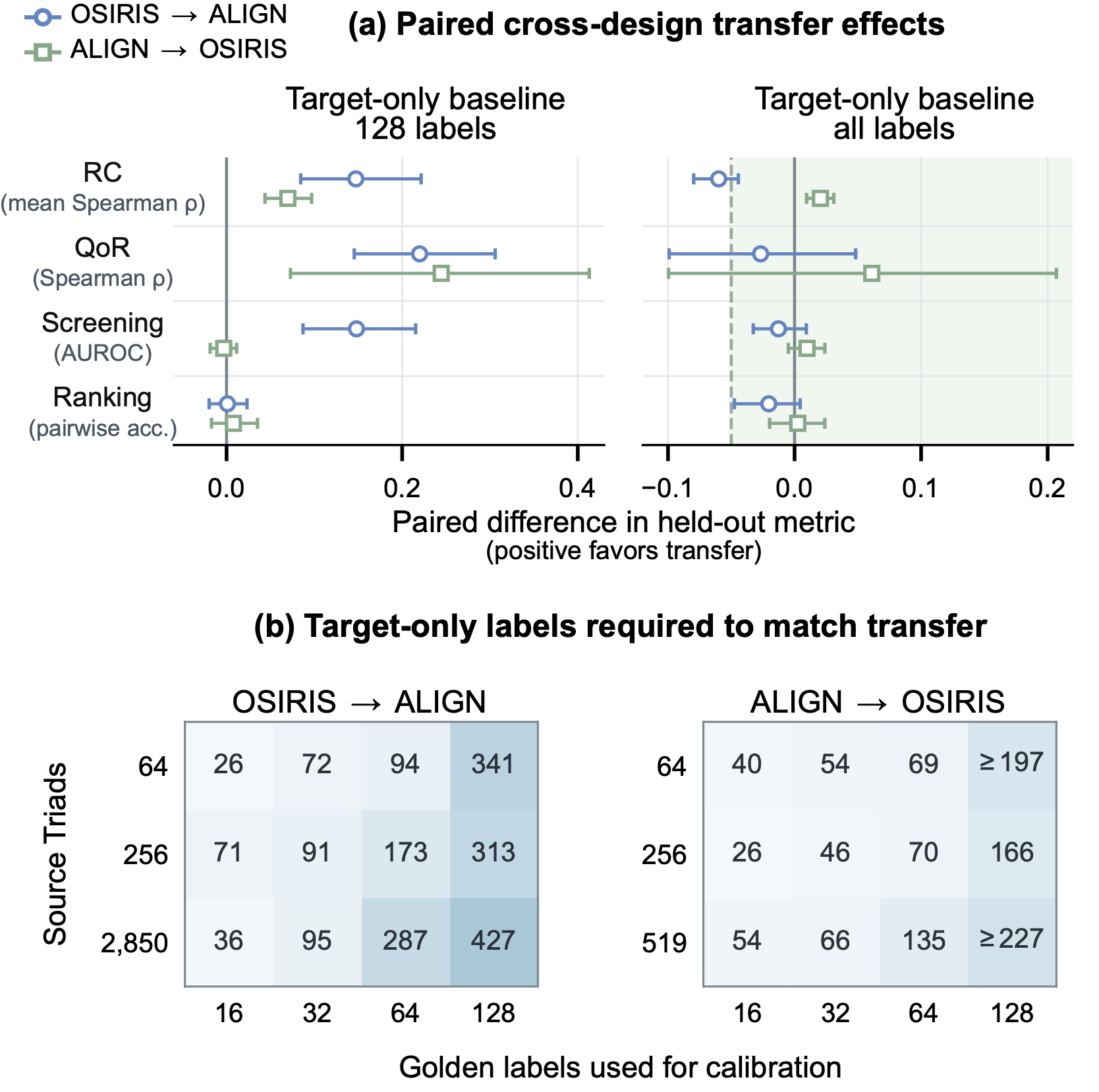}
  \vspace{-6pt}
  \caption{Bidirectional Transfer Comparison.}
  \label{fig:cross-design-transfer}
  \vspace{-15pt}
\end{figure}

\begin{table*}[!t]
  \centering
  \footnotesize
  \setlength{\tabcolsep}{2.2pt}
  \renewcommand{\arraystretch}{1.00}
  
  \caption{Label-efficient L2P calibration versus full-data
  architecture ports on matched downstream tasks.}
  \label{tab:cross-work-native}
  \vspace{-0.85em}

  \resizebox{\textwidth}{!}{%
  \begin{tabular}{@{}l|l|l|ccc|l@{}}
    \hline
    Baseline Port &
    Forward Training Route &
    Matched Task / Metric &
    Full &
    L2P &
    Saved &
    Full $\rightarrow$ L2P Result \\
    \hline

    AutoCkt~\cite{Settaluri2020AutoCkt} &
    Specification/device context $\rightarrow$ MLP QoR surrogate $\rightarrow$ candidate selection &
    5T-OTA Satisfaction $\uparrow$ &
    176 &
    \textbf{100} &
    \textbf{43\%} &
    $75.8\% \rightarrow 91.3\%^{\dagger}$ \\

    GCN-RL~\cite{Wang2020GCNRL} &
    Netlist/device graph $\rightarrow$ GCN metric surrogate $\rightarrow$ maximum-FoM selection &
    Common-Source FoM $\uparrow$ &
    220 &
    \textbf{20} &
    \textbf{91\%} &
    $-0.460 \rightarrow -0.338^{\dagger}$ \\

    CktGNN~\cite{Dong2023CktGNN} &
    Device/subgraph hierarchy $\rightarrow$ hierarchical GNN $\rightarrow$ property regression &
    Cascode Settling RMSE $\downarrow$ &
    256 &
    \textbf{100} &
    \textbf{61\%} &
    $0.648 \rightarrow 0.635$ ns \\

    FALCON~\cite{Mehradfar2025FALCON} &
    Edge-attributed circuit graph $\rightarrow$ message-passing GNN $\rightarrow$ QoR prediction &
    Buffer Timing RMSE $\downarrow$ &
    271 &
    \textbf{50} &
    \textbf{82\%} &
    $8{,}584 \rightarrow 4{,}272$ ps$^{\dagger}$ \\

    \hline
  \end{tabular}%
  }

  \vspace{1pt}
  \parbox{\textwidth}{\scriptsize
  Full is the median outer-training label budget across folds.
  L2P is the earliest budget with a paired 95\% confidence interval excluding
  zero ($^{\dagger}$); CktGNN uses its first point-estimate parity.
  \looseness=-1
  }
  \vspace{-24pt}
\end{table*}

We evaluate four target-domain tasks. \emph{RC prediction} estimates the
log-transformed total extracted capacitance and resistance and is evaluated by
the mean of their two unnormalized Spearman correlations. \emph{QoR regression}
predicts post-layout closed-loop gain in dB for OSIRIS or log-transformed 1\%
settling time for ALIGN, with Spearman correlation measuring whether the
within-target QoR ordering is preserved. \emph{Budget screening} classifies
whether a layout performs better than a fold-specific QoR median computed
exclusively from the non-test target pool; higher gain is preferred for OSIRIS,
whereas lower settling time is preferred for ALIGN. Screening is evaluated by
AUROC. \emph{Candidate ranking} orders valid candidates within the same OSIRIS
circuit family or among layouts derived from the same ALIGN canonical source
schematic and is evaluated by pairwise accuracy. 

Figure~\ref{fig:cross-design-transfer}(a) reports the paired within-task
difference between the transfer and target-only models for every fold and
repeat. The left compares both models at 128 target labels, while
the right uses the all-label target-only baseline. Positive values favor transfer. Figure~\ref{fig:cross-design-transfer}(b) converts performance into target-label equivalents. We invert each task's target-only learning curve at the transfer score
and report the median label count across tasks. At the matched 128-label budget, transfer improves 7 of the 8 task--direction
point estimates, with five confidence intervals excluding zero. 
At the largest pretraining budgets, 128 target labels match 427
target-only labels for OSIRIS$\rightarrow$ALIGN and at least 227 for the
reverse direction, demonstrating substantial label leverage in both directions.

\vspace{-3pt}
\subsection{Cross-PDK Transfer}
\label{sec:cross-pdk-transfer}
\vspace{-2pt}

We evaluate bidirectional process transfer using the SKY130A and GF180MCU-C
standard-cell libraries distributed with Open-PDKs~\cite{OpenPDKs}. Grouping the cells
by canonical Boolean function, we retain 35 functions represented in both
PDKs, comprising 121 SKY130A cells and 125 GF180MCU-C cells. The task structure
follows Section~\ref{sec:cross-design-transfer}, with QoR regression
instantiated separately as Timing and Power.

\begin{figure}[t]
  \centering
  \includegraphics[width=0.85\columnwidth]{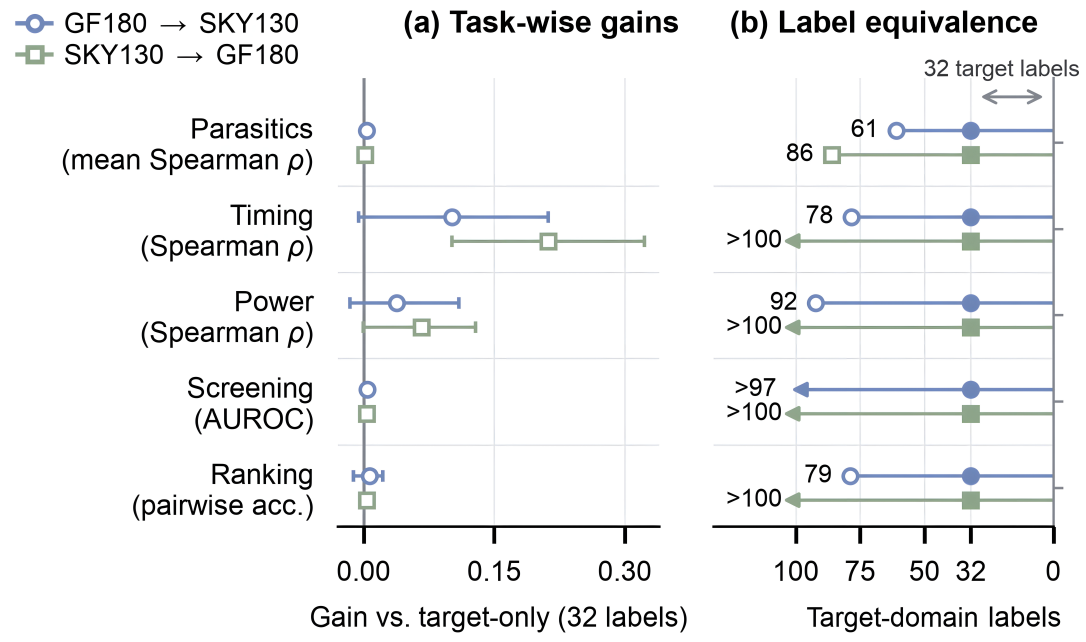}
  \vspace{-8pt}
  \caption{Bidirectional cross-PDK transfer.}
  \label{fig:cross-pdk-transfer}
  \vspace{-12pt}
\end{figure}

Figure~\ref{fig:cross-pdk-transfer}(a) reports paired improvements over
target-only training with the same 32 labels. All ten task--direction point
estimates are nonnegative. The largest gains occur in Timing, reaching $+0.102$
for GF180MCU-C$\rightarrow$SKY130A and $+0.212$ in the
reverse direction; the latter achieves a 95\% confidence interval of
$[0.102,0.322]$. Power also improves by $+0.038$ and $+0.066$, respectively,
while Parasitics, Screening, and Ranking remain slightly above
their corresponding target-only baselines. 

Figure~\ref{fig:cross-pdk-transfer}(b) expresses the same results in the common
unit of target-domain labels. With 32 target labels for power,
GF180MCU-C$\rightarrow$SKY130A reaches the performance of target-only training
with 92 labels, a $2.9\times$ label-equivalence gain.
SKY130A$\rightarrow$GF180MCU-C exceeds the measured target-only endpoint of
100 labels, corresponding to a gain greater than $3.1\times$. This
result provides direct evidence that L2P captures physical relationships that
remain useful across distinct PDK layer and device conventions.

\vspace{-2pt}

\vspace{-3pt}
\subsection{Cross-Work Comparison}
\label{sec:cross-work-comparison}
\vspace{-2pt}

We compare L2P with supervised architecture ports of
AutoCkt~\cite{Settaluri2020AutoCkt},
GCN-RL~\cite{Wang2020GCNRL},
CktGNN~\cite{Dong2023CktGNN}, and
FALCON~\cite{Mehradfar2025FALCON}.
Built from their official implementations, these ports retain
their core input representations and network designs
(Table~\ref{tab:cross-work-native}) and are retrained for QoR
prediction using Huber loss and training-only hyperparameter
selection. The sizing ports replace RL rollouts with
fixed-candidate selection. Both routes score identical held-out
candidates and share training-derived thresholds and normalization,
reward/FoM formulas, argmax selection, and uniform tie handling.
Final satisfaction and FoM are computed from the selected
candidates' true post-layout QoR. CktGNN and FALCON are evaluated
by direct regression on identical held-out records.
All comparisons share source-held-out splits and prediction
targets. Each port uses all outer-training Golden labels,
whereas L2P calibrates a prediction head on its frozen Stage-I
representation. This comparison measures label efficiency
for matched prediction and sizing decisions.
Results aggregate five folds and ten acquisition and calibration
seeds, with confidence intervals bootstrapped over held-out
decision units.
\looseness=-1

\begin{figure}[t]
  \centering
  \includegraphics[width=\columnwidth]{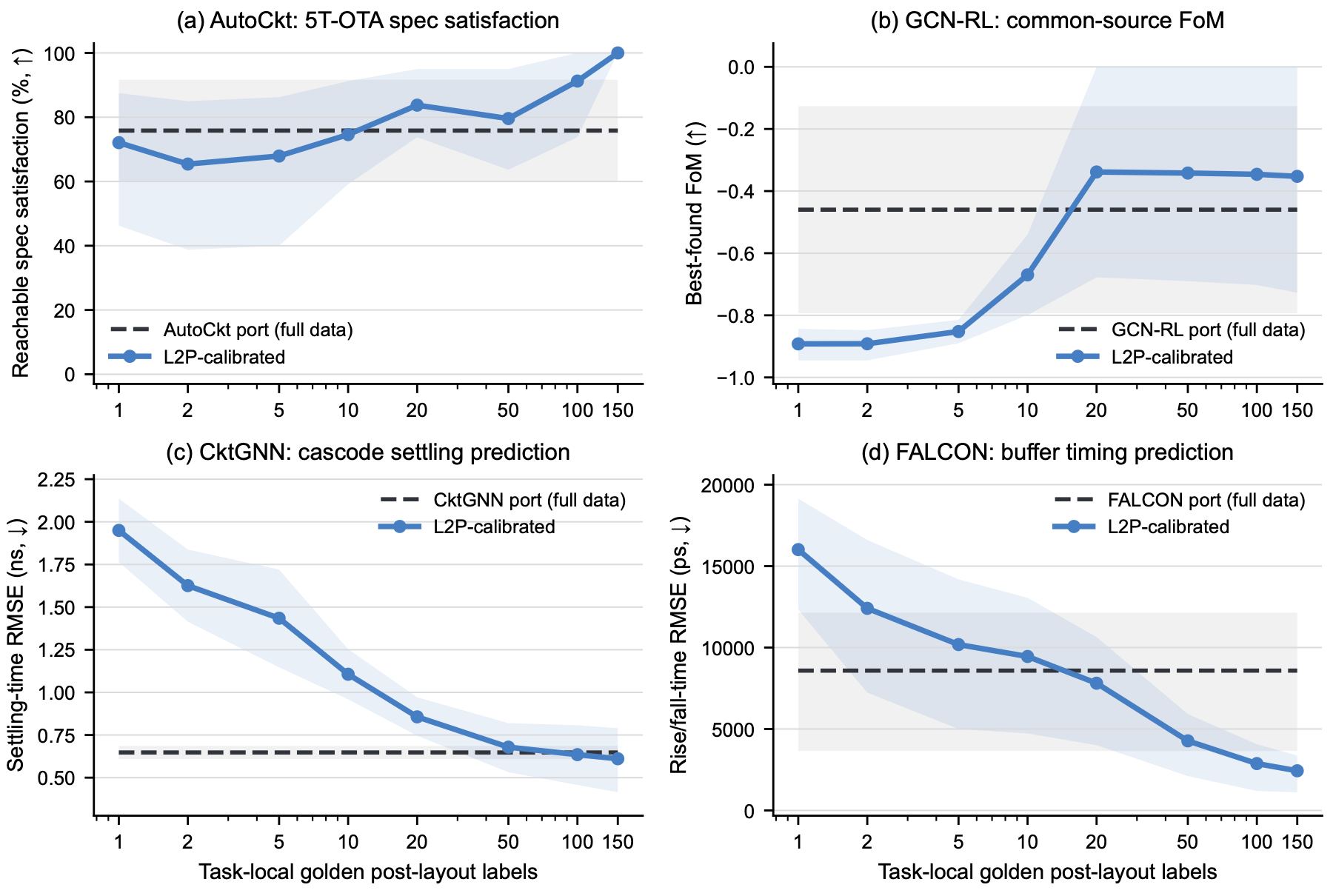}
  \vspace{-15pt}
  \caption{Matched-task performance comparison.}
  \label{fig:paper-native-metrics}
  \vspace{-12pt}
\end{figure}

Figure~\ref{fig:paper-native-metrics} and
Table~\ref{tab:cross-work-native} show that L2P matches or
exceeds all four full-data architecture ports with 20--100
task-local Golden labels. With 100 labels, L2P improves
5T-OTA satisfaction from \(75.8\%\) to \(91.3\%\) and achieves
\(0.635\)-ns cascode settling-time RMSE, compared with
\(0.648\) ns for the corresponding full-data port; with 50 labels, it halves buffer timing RMSE. Paired 95\% confidence intervals exclude zero for improvements in satisfaction, FoM, and buffer timing. Thus, reusable layout-to-physics pretraining
preserves or improves downstream prediction and decision quality
while substantially reducing task-local post-layout supervision.

\section{Conclusion}
\vspace{-1pt}
This work reframes data acquisition for learning-based analog layout automation around the physical evidence encoded in completed
layouts. By decoupling physics from QoR objectives, L2P turns extraction data into a reusable foundation before scarce testbench labels are
introduced. More broadly, L2P suggests that limited task-specific data need not remain a
fundamental barrier to learning-based layout automation. We envision L2P as a step toward automated analog design systems that generalize across circuits, objectives, and processes.

\bibliographystyle{IEEEtran}
\balance
\bibliography{reference}
\end{document}